# What do Support Analysts Know about Their Customers? On the Study and Prediction of Support Ticket Escalations in Large Software Organizations

Lloyd Montgomery and Daniela Damian
*SEGAL Lab, University of Victoria*
Victoria, Canada
lloydrm@uvic.ca, danielad@uvic.ca

*Abstract*—Understanding and keeping the customer happy is a central tenet of requirements engineering. Strategies to gather, analyze, and negotiate requirements are complemented by efforts to manage customer input after products have been deployed. For the latter, support tickets are key in allowing customers to submit their issues, bug reports, and feature requests. Whenever insufficient attention is given to support issues, however, their escalation to management is time-consuming and expensive, especially for large organizations managing hundreds of customers and thousands of support tickets. Our work provides a step towards simplifying the job of support analysts and managers, particularly in predicting the risk of escalating support tickets. In a field study at our large industrial partner, IBM, we used a design science methodology to characterize the support process and data available to IBM analysts in managing escalations. Through iterative cycles of design and evaluation, we translated our understanding of support analysts' expert knowledge of their customers into features of a support ticket model to be implemented into a Machine Learning model to predict support ticket escalations. We trained and evaluated our Machine Learning model on over 2.5 million support tickets and 10,000 escalations, obtaining a recall of 79.9% and an 80.8% reduction in the workload for support analysts looking to identify support tickets at risk of escalation. Further on-site evaluations, through a prototype tool we developed to implement our Machine Learning techniques in practice, showed more efficient weekly support-ticket-management meetings. The features we developed in the Support Ticket Model are designed to serve as a starting place for organizations interested in implementing our model to predict support ticket escalations, and for future researchers to build on to advance research in escalation prediction.

## I. Introduction

Large software organizations handle many customer support issues every day in the form of bug-reports, feature requests, and general misunderstandings as submitted by customers. A significant portion of these issues create new, or relate to, existing technical requirements for product developers, thus allowing requirements management and release planning processes to be reactive to customer input.

These support issues are submitted through various channels such as support forums and product wikis, however, a common default for organizations is to offer direct support through phone and online systems in which support tickets are created and managed by support analysts. The process of addressing these support tickets varies across different organizations, but all of them share a common goal: to resolve the issue brought forth by the customer and keep the customer happy. If a customer is not happy with the support they are receiving, companies have escalation processes whereby customers can state their concern for how their support ticket is being handled by escalating their problems to management's attention.

While the escalation process is needed to draw attention to important and unresolved issues, handling the underlying support ticket after an escalation occurs is very expensive for organizations [1], amounting to millions of dollars each year [2]. Additionally, gathering bottom-up requirements from support tickets is an important requirements-gathering practice for companies looking to address customer feedback and suggestions; however, escalations (and the process of managing them) take time away from support analysts, making the discovery of bottom-up requirements much less efficient. When escalations occur, immediate management and senior software engineers' involvement is necessary to reduce the business and financial loss to the customer. Furthermore, software defect escalations can – if not handled properly – result in a loss of reputation, satisfaction, loyalty, and customers [3].

Understanding the customer is a key factor in keeping them happy and solving support issues. It is the customer who, driven by a perceived ineffective resolution of their issue, escalates tickets to management's attention [4]. A support analyst's job is to assess the risk of support-ticket escalation given the information present – a largely manual process. This information includes the customer, the issue, and interrelated factors such as time of year. Keeping track of customers and their issues becomes infeasible in large organizations who service multiple products across multiple product teams, amounting to large amounts of customer data.

Past research proposed Machine Learning (ML) techniques that model industrial data and predict escalations [1],[2],[4],[5], though none of these efforts attempted to equip ML algorithms with the knowledge-gathering techniques that support analysts use every day to understand their customers. The focus had instead been on improving Escalation Prediction (EP) algorithms while utilizing largely *all available* support data in the studied organization, without much regard to modelling analysts' understanding of whether customers might escalate. Defining which information analysts use to identify issues at risk of escalation is the first step in Feature Engineering (FE): a difficult, expensive, domain-specific task of finding features that correlate with the target class (in this case, escalations) [6]. Using these features in a ML model is designed to leverage the analysts' expert knowledge in assessing and managing the risk of support-ticket escalations to create an automated approach to EP. Additionally, once FE has been completed, these features can serve as a baseline for other organizations with similar processes interested in EP with their own support data.

In our research, we studied this problem in a field study at IBM: a large organization with hundreds of products and customers, and a strong desire to avoid escalations. Two research questions guided our research:

RQ 1. What are the features of a support-ticket model to best describe a customer escalation?
RQ 2. Can ML techniques that implement such a model assist in escalation management?

The contributions of our work have been iteratively developed and evaluated through a design science methodology [7] with our industrial partner, IBM. Our first main contribution is the model of support ticket features – through FE – that support teams use to assess and manage the risk of escalations. This contribution was developed through observations of practice and interviews with management, developers, and support analysts at IBM, as well as analysis of the IBM customer support data repository containing more than 2.5 million support tickets and 10,000 escalations. Our second contribution is the investigation of this model when used with ML techniques to assist in the escalation process. We complemented a statistical validation of our techniques with an in-depth study of the use of these techniques in daily management meetings assessing escalations at one collaborating product team, IBM Victoria in Canada.

## II. RELATED WORK

The development and maintenance of software products is highly coupled with many stakeholders, among which the customer plays a key role. Customer Relationship Management (CRM) involves integrating artifacts, tools, and workflows to successfully initiate, maintain, and (if necessary) terminate customer relationships [8]. Examples of CRM practices include: customer participation requirements-gathering sessions, customer feature suggestions through majority voting, customer incident reports, and support tickets [9],[10]. Other tactics of involving customers in the requirements gathering phase such as stakeholder crowd-sourcing (e.g. Lim et al. [11]) and direct customer participation (e.g. Kabbedijk et al. [9]) are CRM processes that aim to mitigate the potential cost of changing-requirements after development has begun.

An outstanding aspect, however, is the effort and cost associated with the management of a product's ongoing support process: dealing with bugs, defects, and feature requests through artifacts such as product wikis, support chat lines, and support tickets. When support tickets are not handled in a timely manner or a customer's business is seriously impacted, customers escalate their issues to management [2]. Escalation is a process very costly for organizations [2],[4] and yet fruitful for research in ML that can parse large amounts of support ticket data and suggest escalation trends [4],[12].

ML techniques have been proposed in various ways in previous research. Marcu et al. used a three-stage correlation and filter process to match new support issues with existing issues in the system [5]. Their goal and contribution was to speed up the triage and resolution process through finding similar issues previously resolved. Ling et al. [1] and Sheng et al. [2] propose cost-sensitive learning as a technique for improved ML results optimized for EP. Their research, however, was primarily focused on the cost-sensitive learning algorithms and the improvements they offered, with no consideration to the individual attributes being fed into the model. Similarly, Bruckhaus et al. [4] conducted preliminary work investigating the use of neural networks to conduct EP on data from Sun Microsystems. Their work does not describe how they selected their final attributes from an initial set of 200.

The end goal of EP through ML is to identify events generated by customers which might lead to escalations, yet none of the previous research attempts to solve the problem of EP by understanding how analysts identify escalations. Previous research does not focus on the customer through data selection or FE aimed at the knowledge that support analysts have about their customers. Our work addresses this by doing several iterative phases: extensive context-building work within a support organization; iterative cycles of FE focused on understanding the analysts' knowledge of the customer during the support ticket and escalation management process; and finally, real-world deployment of our ML techniques that implement this model to gain feedback on the support ticket features.

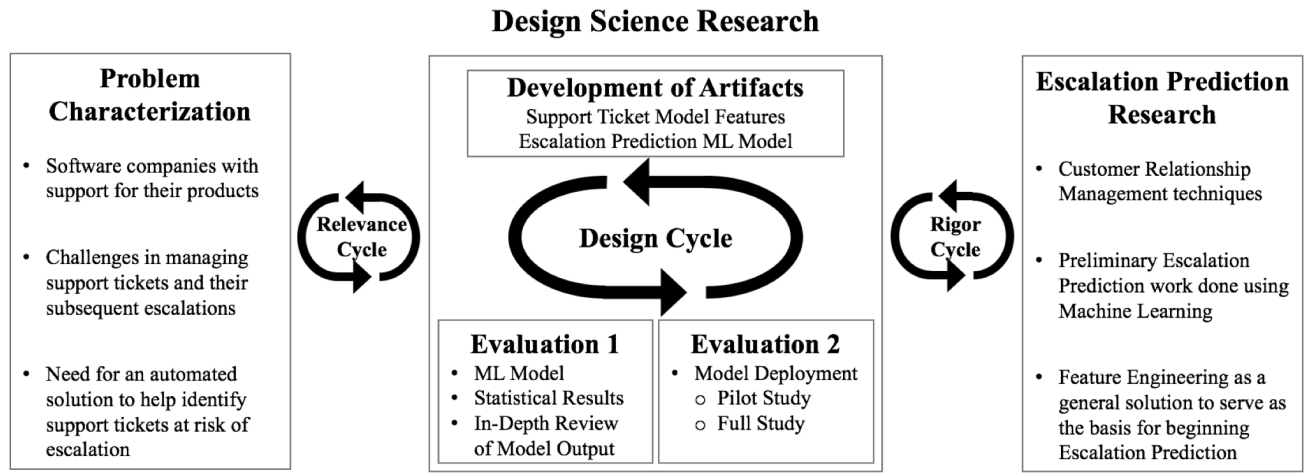

Fig. 1. Design Science research methodology in our study

## III. RESEARCH METHODOLOGY

This research began when IBM approached our research team because of our previous empirical work [13],[14] in investigating development practice in IBM software teams and developing ML solutions to support developer coordination. They described their current problem as: an increasing number of customer issue escalations resulting in additional, costly efforts as well as dissatisfied customers. They sought some automated means to enhance their support process through leveraging the data available in their large customer support repository.

### A. Design Science Approach

To investigate this problem, we employed a design science methodology [7],[15], as illustrated in Fig. 1, whereby artifacts in our research were iteratively developed and evaluated with the stakeholders in the problem domain. First, the design science methodology guided the *problem characterization* phase through various *relevance cycles* including an initial ethnographic exploratory study of the escalation process and data available to IBM customer support analysts. Then, the *development and evaluation of the artifacts* was conducted through multiple *design cycles* with our industry collaborator. Two artifacts were produced: a *Support Ticket Model* of which *features* represent the contextual knowledge held by support analysts about the support process, and the operationalization of those features into an *Escalation Prediction Machine Learning Model*. Both artifacts were iteratively studied and improved through direct support ticket and escalation management activities with our industry collaborator. Finally, to fulfill the *rigor cycle* in our methodology, we reviewed existing work in CRM and EP through ML, and reflected on how our research results are transferrable to other settings.

### B. Study Setting

IBM is a large organization offering a wide range of products to many customers world-wide. In our study, we obtained customer support data consisting of over 2.5 million support tickets and 10,000 escalation artifacts from interactions with 127,000 customers in 152 countries. We also interacted closely with the management and support team at the IBM Victoria site, which employs about 40 people working on two products called IBM Forms and Forms Experience Builder. Several other IBM employees in senior management, worldwide customer support, and Watson Analytics provided us with their input about the support process.

## IV. PROBLEM CHARACTERIZATION

To ground the development of the two artifacts in a deeper understanding of the problem expressed by IBM, we first conducted an ethnographic exploratory study of the IBM support ticket process and escalation management practice. In this section, we discuss the details of our ethnographic study and the insights we obtained towards a detailed characterization of the problem and its context.

### A. Ethnographic Exploratory Study and the Escalation process

To learn about IBM processes, practices, and tools used by support analysts to collect and manage customer support tickets, one of the researchers worked on-site at IBM Victoria for two months. He attended daily support stand-up meetings run jointly by development and support management, and conducted follow-up interviews with management, developers and support analysts. The IBM Victoria staff involved in these sessions included the Victoria Site Manager, the Development Manager, the L3 Support Analyst, and two L2 Support Analysts. Additional information about the IBM support ticket process and escalation management practice was sought through interviews with four other senior analysts and managers at IBM support organizations in North Carolina and California. Additionally, extensive time was spent understanding the data available in the large IBM support ticket repository.

IBM has a standard process for recording and managing customer support issues across all its products. The support process involves multiple levels: L0, ownership verification; L1, basic user-error assistance; L2, product usage assistance from knowledge experts; and L3, development support of bugs and defects. When a new support issue is filed by a customer, a Problem Management Record (PMR) is created to document the lifetime of the issue, including attributes such as severity and priority, and conversations between customers and support. For simplicity, we may use the term *PMR* to refer to a support ticket henceforth in the paper.

IBM handles escalations through a process, and artifact, called a Critical Situation (*CritSit*) that is used when customers are not happy with the progress of their PMR. A PMR is said to "Crit" when a CritSit is opened and that PMR is attached to the CritSit artifact. CritSits can be opened by customers for any reason, although the most likely scenario is to speed up the resolution of their PMR for business or financial reasons. The process of opening and handling a CritSit involves IBM resources in addition to the original resources already being used to solve the issue; furthermore, CritSits are perceived as poor management of PMRs, regardless of the underlying cause. Avoiding and reducing CritSits are top priorities for IBM.

### B. The Problem

Currently, support analysts are tasked with handling PMRs by responding to customer emails: answering questions and offering advice on how to get passed their issue. Manually tracking risk of escalation, however, requires detailed attention beyond the PMR itself, towards the customer behind the PMR: by tracking the business and emotional state of the customer, and ultimately make judgment calls on whether they think a PMR is likely to escalate. This becomes tedious as support analysts manage more and more customers, as each customer within this ecosystem might be related to multiple products and support teams. Dissatisfaction with any of the other products might result in escalations by the customer; furthermore, customer inevitably have trends, repeat issues, and long term historical relationships that might contribute to escalations. To manage the tracking and predictive modelling of all PMRs in the IBM ecosystem, an automated solution is required.

## V. ENGINEERING SUPPORT TICKET MODEL FEATURES

Our approach to addressing the manual process of tracking PMRs and their escalations began by modeling PMR information available to analysts in assessing the possibility of a customer escalating their issue, followed by engineering a set of features into a Support Ticket Model (RQ1). To begin the FE process, we analyzed data from our on-site observations and conducted interviews aimed specifically at understanding how analysts reason through the information about their PMRs and customers. We first describe the interview questions and data we gathered, followed by our data analysis procedure, and then the PMR model features that emerged from our analysis.

### A. Interviews

We conducted a series of semi-structured interviews with support analysts at IBM, five at IBM Victoria and four in worldwide customer support organizations, all of whom are customer facing in their daily jobs. We were interested in identifying information that is currently available in customer records and support tickets, particularly information analysts use to assess the risk of support ticket escalations. We asked questions such as "Why do customers escalate their issues?", "Can you identify certain attributes about the issue, customer, or IBM that may trigger customers to escalate their issue?", as well as exploratory questions about support ticket attributes as we identified in the PMR repository. The full interview script can be found online[1].

### B. Thematic Analysis

We used thematic analysis [16] to analyze our observation notes as well as interview transcripts. From the interviews conducted with IBM, the responses were labelled with feature-like names, thematic *codes*, that represented possible directions for ML features that could automate the process of CritSit prediction. From there we moved on to categories, thematic *themes*, to group the codes based on types of available support- data. The themes and underlying codes are listed in Table I. We validated these themes and codes through two focus groups consisting of: the Victoria Site Manager, the L3 Support Analyst, and an L2 Support Analyst. We then refined the themes and codes from the feedback we received in these meetings.

### C. Support Ticket Model Features

To develop the Support Ticket Model Features we analyzed our customer and support ticket repository data consisting of over 2.5 million PMRs and 10,000 CritSits. We then mapped the PMR attributes to the codes from our analysis under each of the themes we identified. Throughout this process, certain types of PMR data were useable as-is, without modifying the attributes in IBM's dataset such as "number of days open", and other types of data had to be restructured, counted, averaged, or in some cases even engineered from multiple attributes, such as "PMR/CritSit Ratio" which involved two attributes being weighed against each other. Once a code had data mapped to it, it was considered a feature of the model. In developing the model features, we sought to abstract as much as possible from the specifics of IBM's data and processes to increase transferability to other organizations.

TABLE I. PMR-RELATED INFORMATION RELEVANT TO PREDICTING PMR ESCALATIONS

| Themes | Codes |
|---|---|
| IBM Tracked Metrics | How long has a PMR been open |
| Customer Perception of the PMR Process | Fluctuations in severity |
|  | Support analyst involvement |
| Customer Perception of Time with Respect to their PMR | Initial response wait time |
|  | Average response wait time on respective PMRs |
| Traits of Customers | How many PMRs they have owned |
|  | How many CritSits they have owned |
|  | Expectation of response time |

The list of our Support Ticket Model Features is shown in Table II; the list represents the final features as developed through the iterative cycles of our design science methodology. The four feature categories and an initial set of 13 features were created immediately following our thematic analysis, while the additional features (shown in italics in the table) were added as a result of the two evaluation cycles described in Sections VI and VII. We describe each category and the initial 13 associated features below, with explanations from the problem context. The additional features are explained later in the evaluation sections they were engineered from.

**Basic Attributes.** IBM maintains a few useful attributes associated with PMRs for their support analysts to reference. When support analysts are addressing PMRs, the *Number of entries* is a useful attribute that represents how many actions or events have occurred on the PMR to date (e.g. an email is received, a phone call is recorded, the severity increased, etc.). Additionally, the number of *Days open* is a similar attribute that keeps track of days since the PMR was opened.

This feature category, generally lacking in an in-depth analysis of PMRs, is complemented by three other categories that leverage PMR information support analysts identified as most useful in assessing risk of escalation.

**Perception of Process.** Within the support process, there are many people involved with solving customer issues, but there are only a certain *Number of support people in contact with the customer*. If a customer wants to convey the urgency or importance of their issue, the severity attribute on their PMR is the way to do that; customers are in charge of setting the severity of their PMRs. Severity is an attribute from 4 to 1, with 1 being the most severe; severity can be changed to any number at any time. Any *Number of increases in severity* is a sign that the customer believes their issue is becoming more urgent; conversely, any *Number of decreases in severity* can be interpreted as the issue improving. Support analysts watch for increases to severity, but the most severe situations are modelled by the *Number of sev4/sev3/sev2 to sev1* transitions, as this represents the customer bringing maximum attention to their PMR.

---

[1] http://thesegalgroup.org/wp-content/uploads/2017/02/support-analyst.pdf

**Perception of Time.** The customer's perception of time can be engineered using timestamps and ignoring PMR activity that is not visible to the them. The first time when customers may become uneasy is the *Time until first contact* with a support analyst. At this stage the customer is helpless to do anything except wait, which is a unique time in the support process. Once a customer is in contact with support there is an ongoing back-and-forth conversation that takes place through emails and phone calls, the timestamps of which are used to build an *Average support response time*. Each customer has their own expectation of response time, which in turn can be compared to the average response time on the current PMR. This *Difference in average vs expected response time* requires that the customer's expectation of response time is known, which is explained in the next section.

**Customer Profile.** Tracking customer history allows for insights into customer-specific behaviors that manifest as trends across their PMRs. The customer is the gate-keeper of information, the one who sets the pace for the issue, and the sole stakeholder who has anything to gain from escalating their PMR. As such, it seems appropriate to model the customer over the course of all their support tickets. Customers within the IBM ecosystem have a *Number of closed PMRs* and a *Number of closed CritSits*. Combined, these two numbers create a *CritSit to PMR ratio* that represents the historical likelihood that a customer will Crit their future PMRs. Finally, customers have a predisposed *Expectation of support response time* from their past experiences with IBM support. This is calculated by averaging the "Average support response time" feature over all PMRs owned by a customer.

## VI. EVALUATION 1

The next step in our research was to seek validation of the Support Ticket Model Features with IBM support analysts, and to investigate the application of these features in a ML model to predict escalations (RQ2). We evaluated the output of the ML model through statistical validation as well as with IBM support analysts at multiple sites.

*A. Machine Learning Model*

The creation of the ML model was straightforward once PMR data had been mapped to the categories. In total, there were 13 attributes under four categories in the initial set of features that were used in the first stage of training. The model has a binary output as the input of our target class is 0 or 1. Most models, including the one we selected, output a confidence in that prediction, and we chose to correlate that to Escalation Risk (ER). For example, if the model output a prediction of 1, with confidence 0.88, this PMR's ER is 88%. Any ER over 50% is categorized as a Crit in the output of the model.

TABLE II. SUPPORT TICKET MODEL FEATURES

| Feature | Description |
|---|---|
| **Basic Attributes** | |
| Number of entries | Number of events/actions on the PMR |
| Days open | Days from open to close (or CritSit) |
| *Escalation type* | *CritSit Cause, CritSit Cascade, or None* |
| *PMR ownership level* | *Level of Support (L0 – L3) that is in charge of the PMR, calculated per entry* |
| **Perception of Process** | |
| Number of support people in contact with customer | Number of support people the customer is currently communicating with |
| Number of increases in severity | Number of times the Severity increase |
| Number of decreases in severity | Number of times the Severity decrease |
| Number of sev4/sev3/sev2 to sev1 transitions | Number of changes in Severity from 4/3/2 to 1 |
| **Perception of Time** | |
| Time until first contact | Minutes before the customer hears from IBM for the first time on this PMR |
| Average support response time | Average number of minutes of all the support response times on this PMR |
| Difference in average vs expected response time | (Expectation of support response time) minus (Average support response time) |
| *Days since last contact* | *Number of days since last contact, calculated per entry* |
| **Customer Profile** | |
| Number of closed PMRs | Number of PMRs owned by customer that are now closed |
| Number of closed CritSits | Number of CritSits owned by customer that are now closed |
| CritSit to PMR ratio | (Number of CritSits) over (Number of PMRs) |
| Expectation of support response time | Average of all "Average support response time" of all PMRs owned by a customer |
| *Number of open PMRs* | *Number of PMRs this customer has open* |
| *Number of PMRs opened in the last X months* | *Number of PMRs this customer opened in the last X months* |
| *Number of PMRs closed in the last X months* | *Number of PMRs this customer closed in the last X months* |
| *Number of open CritSits* | *Number of CritSits this customer has open* |
| *Number of CritSits opened in the last X months* | *Number of CritSits this customer opened in the last X months* |
| *Number of CritSits closed in the last X months* | *Number of CritSits this customer closed in the last X months* |
| *Expected support response time given the last X months* | *Average of all "Average support response time" of all PMRs owned by a customer in the last X months* |

We fed the 13 original Support Ticket Model Features into multiple supervised ML algorithms: CHAID [17], SVM [18], Logistic Regression [19], and Random Forest [18]. Although other algorithms produced higher precision, we chose Random Forest because it produced the highest recall. High recall was preferred for two reasons: as argued by Berry [20] and exemplified in the recent work of Merten et al. [10]. Additionally, our industrial partner expressed a business goal of identifying problematic PMRs while missing as few as possible. The input we received from the IBM analysts was that they would prefer to give more attention to PMRs that have potential to Crit, rather than potentially missing CritSits.

The ratio of CritSit to non-CritSit PMRs is extremely unbalanced at 1:250, therefore some kind of balancing was required to perform the ML task. The Random Forest classifier we used has the capability to handle imbalanced data using oversampling of the minority class [18]. In other words, the algorithm re-samples the minority class (CritSit) roughly enough times to make the ratio 1:1, which ultimately means that each of the minority class items are used 250 times during the training phase of the model. This method allows all 2.5 million of the majority class items to be used in learning about the majority class, at the cost of over-using the minority items during the learning phase.

### B. Statistical Results & Validation

The 2.5 million PMRs and 10,000 CritSits were randomly distributed into 10 folds, and then 10-fold leave-one-out cross-validation was performed on the dataset using the Random Forest classifier. The results of the validation can be seen in the confusion matrix in Table III. A confusion matrix is a useful method of analyzing classification results [21] that graphs the True Positives (TP), True Negatives (TN), False Positives (FP), and False Negatives (FN). The diagonal cells from top-left to bottom-right represent correct predictions (TN and TP).

The recall for "CritSit – Yes" is 79.94%, with a precision of 1.65%. Recall and precision are calculated as $\frac{TP}{TP+FN}$ and $\frac{TP}{TP+FP}$, respectively. The recall of 79.94% means that the model is retrieving 79.94% of the relevant PMRs (CritSits), whereas the precision of 1.65% means that the algorithm is retrieving a lot more Non-CritSit PMRs than CritSit PMRs, so much so that the ratio of CritSit PMRs to all PMRs retrieved is 1.65%.

As previously mentioned, our business goal for building the predictive model was to maximize the recall. Additionally, Berry et al. [22] argue about tuning models to predict in favor of recall when it is generally easier to correct FPs than it is to correct TNs. Significant work has been completed towards identifying which of the PMRs are CritSits, this work is measured through the metric "summarization", calculated as such:

$$\frac{TN + FN}{TN + FN + TP + FP}$$

In short, summarization is the percentage of work done by classification algorithms towards reducing the size of the original set, given that the new set is the sum of FP + TP [20]. Summarization alone, however, is not useful, it must be balanced against recall. 100% recall and any summarization value greater than 0% is progress towards solving identification and classification problems. Our model has 79.94% recall and 80.77% summarization. Simply put, if a support analyst wanted to spend time identifying potential CritSits from PMRs, our model reduces the number of candidate PMRs by 80.77%, with the statistical guarantee that 79.94% of CritSits remain.

### C. Model Output Validation

Using our close relationship with IBM Victoria, we then conducted an in-depth review of the model output in a 2-hour meeting with the support analysts and managers, to gain deeper insights into the behavior of the model on an individual PMR-level basis, to improve the model features.

TABLE III. CRITSIT PREDICTION CONFUSION MATRIX

| Actual | Total | Predicted as | |
|---|---|---|---|
| | | *CritSit - No* | *CritSit - Yes* |
| *CritSit - No* | 2,557,730 | 2,072,496 (TN) 81.03% | 485,234 (FP) 18.97 % |
| *CritSit - Yes* | 10,199 | 2,046 (FN) 20.06% | 8,153 (TP) 79.94 % |

*1) Study Setting*

We examined ten major (suggested by IBM) closed CritSit PMRs from IBM Victoria in our dataset and ran our ML model to produce escalation-risk graphs for each of the CritSit PMRs. The ten CritSit PMRs chosen by IBM were memorable escalations, memorable enough to be discussed with clarity. We show six of the ten graphs in Fig. 2-4, each graph is a single PMR. The graphs plot the ER as produced by our ML model over time, from the first snapshot to its last snapshot. By "snapshot" we are referring to the historical entries that exist per PMR. E.g., a PMR with 16 changes to its data will have 16 snapshots, each consecutive snapshot containing the data from the last snapshot plus one more change. Our goal was to compare the output of our model with what IBM remembered about these ten PMRs when they were handled as escalating issues (i.e. at the time of each snapshot).

The 2-hour in-depth review involved four IBM support representatives: the Site Manager, the Dev. Manager, the L3 Support Analyst, and an L2 Support Analyst. We printed the graphs of these ten CritSit PMRs, discussed them as described below, and took notes during the meeting:

a. Revealing to the members PMR numbers and customer names of the PMRs in the analysis, allowing them to look up these PMRs in their system and read through them.
b. Discussed the PMRs in the order the members preferred.
c. Displayed the graphs of the Escalation Risks.
d. Inquired about how the model performed during each PMR in comparison to what they experienced at the time.

*2) Study Findings*

Overall, our ML model performed well in predicting the ER per PMR, per snapshot. However, the findings of this in-depth review of the model are broader and pertain to a) improvements in our model with respect to the Customer Profile information and b) our increased understanding of IBM's support process. Both findings relate to refinements in our model as well as recommendations to other organizations intending to apply our model to perform EP.

*a) Role of Historical Customer Profile Information*

Two of the ten PMRs in this evaluation showed a trend of building ER over time as events occurred, as shown in Fig. 2. Manual inspection and discussion with the analysts indicate that this behavior was correlated with a lack of Customer Profile information for both PMRs. All Customer Profile features (see Table II) refer to data that is available when the PMR is created and will not change during the lifetime of the PMR; therefore, the initial ER is solely due to the Customer Profile features, and the changes in ER during the lifetime of the PMR must be due to the other three categories.

In contrast, PMRs with too much Customer Profile information were immediately flagged as CritSits. The model had learned that excessive Customer Profile information correlates with high ER. Five of the ten PMRs had this behavior, two of which can be seen in Fig. 3. Manual inspection of the five PMRs revealed a lot of Customer Profile information for each of the five PMRs, i.e., the "Number of Closed PMRs" field was 200+ for each of the five customers of these PMRs.

These findings show variance in model performance for the two extremes of quantity of Customer Profile information in the PMRs we studied. We saw expected behavior for lack of Customer Profile information but unexpected behavior for the opposite, PMRs with extensive Customer Profile information. These variances point to the role of the Customer Profile category in capturing aspects of the customer beyond the current PMR, allowing traits of the customer to be considered during the prediction of escalation risk. To properly capture the features of the Customer Profile category, we made refinements to our model by adding new attributes that add decay of customer information over time, such that the history does not exist forever. These attributes, indicated in italics in Table II, are: "*Number of PMRs opened in the last X months*" and "*Number of CritSits opened in the last X months*" as well as the revised attributes "*Number of PMRs Closed in the last X months*", "*Number of CritSits closed in the last X months*", and "*Expected support response time given the last X months*". We also added features to represent the current state of the customer's involvement with the support team: "*Number of open PMRs*" and "*Number of open CritSits*".

*b) Recording True Reason for CritSit PMRs is Important*

The second insight from this study was about IBM's support process and feedback into revised features in our model. We ran into a situation where on some of the PMRs our model showed low ERs, although they appeared officially as CritSits in the IBM system. We discovered that it is common practice to Crit every PMR owned by a customer when any one of their PMRs Crit. Therefore, there was a distinction between the "cause" CritSit – the CritSit PMR that caused the Crit to happen, and "cascade" CritSits – the CritSit PMR(s) that subsequently Crit due to the process of applying a Crit to every PMR owned by the same Customer in response to some "cause" CritSit. Figure 4 shows two of the three PMRs that had this behavior ("cascade" CritSits) in which our model behaved correctly.

Through manual inspection of PMR historical information, our study participants identified that these three PMRs were not the cause of the CritSit, and in fact there were other PMRs with the same CritSit ID that were responsible for them being recorded as CritSits in the IBM system. Therefore, we recommended to IBM to track the difference between "cause" and "cascade" CritSits for a proper separation of the data. We also added a new feature to our model, "*Escalation Type*".

## VII. EVALUATION 2

The second evaluation investigated the assistance provided by our model running in real time during the management meetings at the Victoria site when analysts together with management

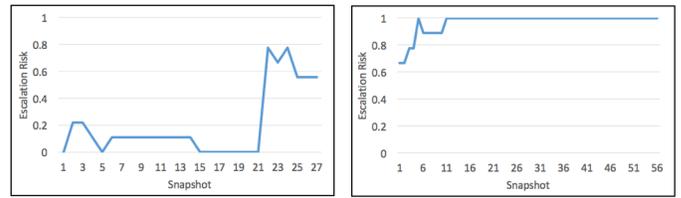
Fig. 2. PMRs with little-to-no Customer Profile info build ER over time

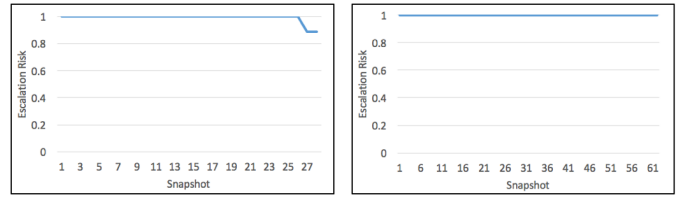
Fig. 3. PMRs with too much Customer Profile info default to high ER early

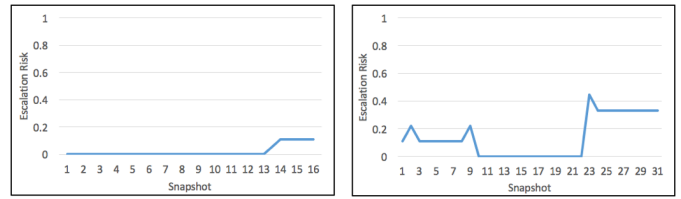
Fig. 4. "Cascade" CritSits showed low ER

discussed open PMRs. To do this, we developed a prototype tool that displays all open PMRs and their current predicted ER, as well as the 13 calculated features – per PMR – that go into the prediction.

### A. Our Prototype

Our prototype tool displayed all active PMRs at the Victoria site with two main displays: the overview, and the in-depth view. The overview displays all open PMRs in a summarized fashion for quick review. The in-depth view comes up when a PMR is selected and shows the details of the PMR. Included in this view is: the history of email correspondence between support and customer, description of the issue, and the ML model features that were used to produce the escalation risk (ER).

### B. Study Setting

We evaluated the use of our prototype over a period of four weeks during daily stand-up support meetings with managers and support analysts. The only tool being used to track PMRs day-to-day before our study was an excel sheet stored locally on the Site Manager's computer. The effectiveness of the meetings relied on support analysts to bring up and discuss PMRs they were working on.

Our prototype was first integrated in a pilot study, to gain feedback on shortfalls and bugs. After the short (one week) pilot, a week was spent improving the tool based on recommendations before the full four-week deployment. The participants of this study were the Victoria Site Manager, the Development Manager, the L3 Support Analyst, and two L2 Support Analysts. One of the researchers participated in all these meetings while the tool was in use for the first two weeks of the study, as well as two days near the end of the study.

After the pilot study two additional features were added to the tool: (1) Displaying a Manual Escalation Risk (MER), a number field from 0 to 100 (to be input by anyone on the team) to eliminate the need to remember the analysts' assessments of each PMR during past meetings; and (2) Displaying a Change in Escalation Risk (CER), a number field from -100 to 100 that represents the change in ER since the last update, to eliminate the need for anyone to memorize ERs by tracking changes manually. With the MER and CER being tracked and displayed, the team could expedite the daily PMR review process and focus on PMRs that either had a high MER or CER.

*C. Study Findings*

The use of our prototype during the PMR management meetings increased their efficiency. In the absence of our tool, the analysts would review PMRs brought up by support analysts and discuss them based on the memory of the participants, often relying on management to bring up additional items they had forgotten. With our tool, they were able to parse through a list of PMRs ranked by ER. The MER capability allowed them to record their own assessment of the ER, and compare it with the ER output by our ML model. It allowed for subsequent meetings to be quicker because the team could see their past evaluations of PMRs, and focus on ones they had assigned a high MER. The CER field provided a quick reference to which PMRs had increased in ER since the last update.

During the study, we observed that a high risk of escalation was often correlated to the same types of customer problems. The team also identified that there were two important aspects of PMRs that mattered to them as well as the customer: PMR ownership level, and days since last contact. PMRs are always being directly managed by some level of support, and the difference between L2 and L3 support means a lot to IBM as well as the customer. L2 is product-usage support, where customers are generally at fault, and L3 is development-level support, where bugs are triaged and the product is at fault. Similarly, the number of days since last customer contact was brought up as an important factor for deciding when a customer may Crit. As a result of these discussions, two new features were added to our final set of model features in Table II: "*PMR ownership level*" and "*Days since last contact*".

## VIII. DISCUSSION

Prompted by the problem of inefficiency in managing customer support ticket escalations at our industrial partner IBM, our approach had been to study and model the information available to support analysts in assessing whether customers would escalate on a particular problem they reported, and to investigate ML techniques to apply this model to support the escalation management process. We employed a design science methodology and here we discuss, as outlined by Sedlmair et al. [7], our contributions through three main design science aspects: problem characterization and abstraction, validated design, and reflection.

*A. Problem Characterization*

The investigation of IBM support practices in our ethnographic study was the first step in our design science iterative process, providing a more detailed understanding of the support ticket escalation problem at IBM. We elaborate here on two lessons learned during the problem characterization phase.

The first lesson we learned is about the importance of this step and iterating through it in the design study. From our initial interviews with the support analysts we were able to draw an understanding of how they work as well as the initial list of our PMR model features. However, it was only after the first evaluation step (the in-depth investigation of the ten CritSit PMRs at the Victoria site) that we reflected and refined our understanding of the problem context in the analysts' job. We were able to uncover details of the cascading CritSits process and its effect on how data was being presented to the analysts. This turned out to be crucial to understanding the PMR life-cycle and to refinements in our PMR model features.

The second lesson relates to abstracting from the specifics of IBM relative to data that can be modeled for EP in other organizations. We learned that some elements of the support process may be intentionally hidden from customers to simplify the support process for them, but also to protect the organization's information and processes. An example of this is the offline conversations that occur between people working to solve support tickets: a necessary process of information sharing and problem solving, but these conversations are never revealed to customers. Other organizations might have similar practices, and being aware of the distinction between customer-facing and hidden information is important. We recommend that companies experiment with both including and not including information hidden from customers in their ML models. Information not known to their customers may be introducing noise to their models.

*B. Validated Support Ticket Model Features*

The two artifacts we iteratively developed in our design science methodology are the Support Ticket Model Features and their implementation into an EP ML model to assist support analysts in managing support-ticket escalations. We believe that the major, unique contribution of this research is the Support Ticket Model. Its features were not only derived from an understanding of support analysts at our industrial partner, but were iteratively refined through several validations of the EP ML techniques that implemented it.

The task of predicting support-ticket escalations is fundamentally about understanding the customers' experience within the support ticket management process. The features we created in our model were designed to represent the knowledge that support analysts typically have about their customers. Through the process of FE, our work identified the subset of features relevant to EP from an understanding of practice around escalation management. Finally, we sought to abstract from IBM practice towards a general model of the escalation management process, and therefore have our results be applicable to support teams in other organizations.

Once the Support Ticket Model Features had been created, they were used in a ML model, the Random Forest classifier. The results of the 10-fold cross validation (shown in Table III) were promising, with a recall of 79.94% and summarization of 80.77%. Our collaborating IBM support team was very pleased with this result, as an 80.77% reduction in the workload to identify high-risk PMRs is a promising start to addressing the reduction of CritSits.

Finally, a prototype tool was built to integrate the real-time results of putting live PMRs through our model to produce escalation risks. Use of our prototype tool granted shorter meetings addressing more issues focused on support tickets deemed important by IBM and the ML model, while still allowing for longer meetings to review more PMRs if they needed to. The main benefit was the summarization and visualization of the support tickets based on a combination of our model output as well as their own assessment through the MER.

*C. Reflection*

Our work adds to the scarce research into automating the prediction of support ticket escalations in software organizations. We reflect below on the relationship between our work and these existing techniques, and discuss implications for practitioners who wish to use this work.

*1) Limitations in Addressing Previous Research*

The work done by both Ling and Sheng and colleagues [1], [2] involves improvements to existing ML algorithms using cost-sensitive learning algorithms, with no consideration to the attributes being fed into the model. The option of using their work as a baseline to compare precision and recall required our data to be in such a format that it could be run through their algorithms. Our data, however, was not fit for classification-based ML algorithms because it is archival, with multiple historical entries per each support ticket. Basic classification ML algorithms require there to be one entry per support ticket, so any archival data such as ours would have to go through a process to convert that data into a summarized format. The final summarized data depends on the conversion process chosen; therefore, we could not simply convert our data and hope it conformed to the constraints of the previous studies due to the lack of information regarding their data structures.

The work done by Bruckhaus et al. [4] has a similar data processing issue, except their work involved some FE to convert attributes into a usable form. They neither describe how they conducted their FE nor the final set of engineered features, therefore we could not compare FE results. Furthermore, the details about their neural network approach, including the parameters and tweaks made to their proposed algorithm, are not provided, making its replication very difficult.

Given the lack of ability to replicate the process and results of previous work with our data, we were not able to contrast our work against this related work; instead, our research focused on FE and iteratively developing our predictive model with support analysts through our design-science approach.

*2) New Directions for Further Improving the Model*

Our work represents a first step towards a model of support ticket information through FE relevant to predicting the risk of support ticket escalations; however, further validation of our model (with its complete set of features) is needed. Through our design-science iterative cycles, we discovered improvements for the model features but we were not able to include them all into the ML implementation due to limitations of our available data.

These improvements inform new research questions that would allow further development of the model, for example:
- What is a meaningful time window for the decay of customer history? (One month, six months, etc.)
- What features would better represent customers within organizations? (Open tickets, number of products owned, etc.)
- Would certain subsets of the data (countries, product areas, product teams, etc.) increase the precision?
- Would sentiment analysis on conversations with the customer during the escalation process improve the model?
- Could NLP techniques be employed to automatically classify the types of customer problems and would certain type of problems correlate with high risk of escalations?
- Is there a business impact by using this model and its supporting tools? Are there economic savings?

*3) Implications for Practitioners*

The model we developed has the potential for deployment in other organizations given that they have enough available data and the ability to map it to the features provided by our model. To implement the ML-based EP model we developed, organizations must track, map, and augment their data to the Support Ticket Model Features. If the high recall and summarization we obtained at IBM is obtained at other organizations, there is potential to reduce their escalation identification workload by ~80%, with the potential for ~80% of the escalations to remain in the reduced set. If this frees up time for support analysts, then they can put additional effort into more important aspects of the support process like solving difficult issues and identifying bottom-up requirements from support tickets.

Prior to implementing our model, organizations should do a cost-benefit analysis to see if the potential benefits are worth the implementation effort. Included in this analysis should be the cost of a support ticket – with and without an escalation, as well as time required to manually investigate tickets, customers, and products for escalation patterns. If the overall cost of escalating tickets and the investigative efforts to avoid escalations outweigh the overall time-spent implementing the model described above, then there is a strong case for implementation.

IX. THREATS TO VALIDITY

The first threat, to *external validity* [23], is the potential lack of generalizability of the results due to our research being conducted in close collaboration with only one organization. To mitigate this threat, the categories and features in our support ticket model were created with an effort of abstracting away from any specifics to IBM processes, towards data available and customer support processes in other organizations.

The second threat, to *construct validity* [23], applies to the mapping of the information and data we collected through interviews with support analysts to the thematic themes and codes. To mitigate that threat, multiple techniques were used: member checking, triangulation, and prolonged contact with participants [23]. The design science method of iteratively working with industry through design cycles puts a strong emphasis on member checking, which Lincoln and Guba [24] describe as "the most crucial technique for establishing credibility" in a study with industry. We described our themes and codes to our IBM analysts and managers, to validate that our data mappings resonated with their practice, through focus groups and general discussions about our results. Triangulation, through contacting multiple IBM support analysts at different sites as well as observations of their practice during support meetings, was used to search for convergence from different sources to further validate the features and mappings created [25]. Finally, our contact with IBM during this research lasted over a year, facilitating prolonged contact with participants which allowed validation of information and results in different temporal contexts.

The third threat, to *internal validity* [23], relates to the noise in the data discovered during the iterative cycles of our design science methodology. As discussed in Section IV, the CritSits in our dataset could be "cause" or "cascade". Due to limitations of our data, we are unable to reliably tell the two types of CritSits apart; however, there is a small subset of CritSits we know for sure are "cause" CritSits. At the cost of discarding many "cause" and uncertain CritSits, we removed all "cascade" CritSit PMRs by discarding the CritSits that had more than one associated PMR. The newer "real" CritSit PMRs (CritSits with only one PMR attached) in our data then totaled ~3,500 (35% of our original target set). The recall on the new target set was 74.47%, with a summarization of 82.85%, meaning that the threat to internal validity due to this noise in our data was negligible.

## X. Conclusion

Effectively managing customer relationships through handling support issues on ongoing software projects is key to an organization's success, and one practice that informs activities of requirements management. Support analysts are a key stakeholder in gathering bottom-up requirements, and proper management of support ticket escalations can allow them to do their job with less attention to escalations. The two artifacts we developed in this work, the Support Ticket Model Features and its implementation in a ML classifier to predict the risk of support ticket escalation, represent a first step towards simplifying support analysts' job and helping organizations manage their customer relationships effectively. We hope that this research leads to future implementations in additional industry settings, and further improvements to EP through ML in future research.

## Acknowledgments

We thank IBM for their data, advice, and time spent as a collaborator; special thanks to Keith Mackenzie at IBM Victoria for his contribution to this research. We thank Emma Reading for her contribution to the prototype tool. This research was funded by NSERC and IBM Center for Advanced Studies.